\documentclass[prd,%
preprint, %
groupedaddress,showpacs,nofootinbib, %
eqsecnum,amsfonts,amsmath,amssymb
              ]{revtex4}

\usepackage{bm}
\usepackage{graphicx}
\usepackage[mathscr]{euscript}

\newcommand{\R}{\mathscr{R}}
\newcommand{\F}{\mathscr{F}}
\newcommand{\A}{\mathscr{A}}
\newcommand{\Rs}{\mathscr{S}}
\newcommand{\Hom}{\mathscr{H}}

\allowdisplaybreaks

\begin{document}

\title{On the structure of the post-Newtonian expansion \\in general
relativity}

\date{\today}

\author{Luc Blanchet}
\email{blanchet@iap.fr}

\author{Guillaume Faye}
\email{faye@iap.fr}

\author{Samaya Nissanke}
\email{nissanke@iap.fr}

\affiliation{${\mathcal{G}}{\mathbb{R}}\varepsilon{\mathbb{C}}{\mathcal{O}}$
-- Gravitation et Cosmologie,\\ Institut d'Astrophysique de Paris, C.N.R.S.,\\
98$^{\text{bis}}$ boulevard Arago, 75014 Paris, France}

\date{\today}

\begin{abstract}
In the continuation of a preceding work, we derive a new expression for the
metric in the near zone of an isolated matter system in post-Newtonian
approximations of general relativity. The post-Newtonian metric, a solution of
the field equations in harmonic coordinates, is formally valid up to any
order, and is cast in the form of a particular solution of the wave equation,
plus a specific homogeneous solution which ensures the asymptotic matching to
the multipolar expansion of the gravitational field in the exterior of the
system. The new form is suitable for practical computations of the
gravitational radiation reaction forces in harmonic coordinates. It also
provides some insights on the structure of the post-Newtonian expansion in
general relativity and the radiation reaction terms therein.
\end{abstract}

\maketitle

\section{Introduction} \label{sec:introduction}

In a previous paper \cite{NB05} (hereafter referred to as paper I), the
equations of motion of compact binary systems were computed at the 3.5
post-Newtonian (PN) order, corresponding to the formal $ 1/c^7 $ level in a
post-Newtonian expansion with respect to the Newtonian acceleration (where $ c
$ is the speed of light). This confirmed and extended previous calculations in
Refs.~\cite{IW93, JaraS97, PW02, KFS03}. The 3.5PN terms consist of
gravitational radiation reaction terms, 1PN order beyond the lowest order of
radiation reaction, which occurs at 2.5PN order in the acceleration. To deal
with the divergences of Poisson or Poisson-like integrals in higher
orders (a well-known problem in the standard post-Newtonian
formalism), paper I used a particular solution of the Poisson equation
defined by means of a finite part operation resorting to complex
analytic continuation.

The aim of the present paper is to justify fully the procedure that was
employed in paper I to compute high-order post-Newtonian terms (in particular
the radiation reaction terms therein), and that was based on a formal
expansion of the retardation of the standard retarded integral, augmented
by the finite part operation. The proof will essentially consist of deriving a
new expression for the post-Newtonian expansion in the near zone of a general
isolated post-Newtonian source, which will constitute an interesting variant
of the solution proposed in a preceding article \cite{PB02} (hereafter paper
II). The latter solution is obtained within a particular framework combining
the multipolar-post-Minkowskian formalism with a matching to a post-Newtonian
source \cite{BD86, B95, B98mult}. 

More precisely, in paper II, a general expression is provided for the
gravitational field generated by an isolated (slowly-moving) system in its
near zone, developed up to any post-Newtonian order. This expression is free
of the well-known divergences of Poisson-type integrals which had plagued the
post-Newtonian expansion at high-order approximations in earlier works (see
\textit{e.g.} Refs.~\cite{PB59, CN69, CE70, ADec75, Ehl80, Kerlick80a,
Kerlick80b, PapaL81} for discussions). The method adopted in paper II to deal
with these divergences has been to replace the Poisson integrals by a new type
of solution to the hierarchy of post-Newtonian equations. It involves a
``regularization'' of the boundary of the Poisson integral at infinity by
means of the finite part. Let us emphasize that the finite part is only used
as a mathematical trick to find in a convenient way a \emph{particular}
solution of the differential equation we want to solve. To this particular
solution, we add a \emph{homogeneous} solution which is determined by
asymptotic matching to the external field. It is then possible to iterate
formally the post-Newtonian expansion \textit{ad infinitum}.

The matching must be understood as a variant of the method
of asymptotic-expansion matching applicable in the exterior part of the
source's near zone \cite{Fock, Bu71, BD88, BD89, B98mult}. As the exterior
solution satisfies a condition of no-incoming radiation, the solution of paper
II incorporates the effects of radiation reaction and, therefore, gives a
correct physical description of the isolated system.

The alternative form that we derive for the end result of paper II further
clarifies its relationship with the usual method of computing the successive
post-Newtonian approximations, based on a naive expansion of the field
equations, like for instance in Ref.~\cite{ADec75}. Essentially, we show that
the present method leads to correct results up to the 3.5PN order; this is,
however, no longer the case starting from the 4PN order $ \sim 1/c^8 $,
because of the non-linear contributions of gravitational wave tails
\cite{BD88, B93, B97}. As we shall see, the expression of the post-Newtonian
expansion derived below provides further insights on the structure of the
gravitational radiation reaction terms and the contribution of tails in the
near zone.

Section \ref{sec:reminder} recalls some necessary results from paper II, and
offers complementary comments. In Section \ref{sec:new_PN_expansion}, we
derive the novel expression for the post-Newtonian expansion, which we
subsequently discuss. Finally, in Section \ref{sec:harmonicity}, we check
directly, by using our new form, that this solution is divergenceless, and
thus satisfies the harmonic coordinate Einstein field equations. The paper
ends in Section \ref{sec:conclusion} with a short conclusion.

\section{Reminders of earlier work and comments} \label{sec:reminder} 

In the following, we shall describe the gravitational field in harmonic
coordinates, assumed to exist globally and to be unique, in terms of the
``Gothic'' metric deviation from flat space-time, $ h^{\mu\nu} $. We shall
thus have
\begin{subequations}\label{eq:harmonicity}\begin{align}
& \partial_\nu h^{\mu\nu} = 0 , \\ & h^{\mu\nu} \equiv
\sqrt{-g}\,g^{\mu\nu}-\eta^{\mu\nu} ,
\end{align}
\end{subequations}%
where $ g^{\mu\nu} $ and $ g $ are the inverse matrix and the determinant of
the usual covariant metric $ g_{\alpha\beta} $ respectively, and where $
\eta^{\mu\nu} = \mathrm{diag}(-1,1,1,1) $ denotes the Minkowski flat metric in
Cartesian coordinates. With this choice and notation, the ``relaxed'' Einstein
field equations simply read
\begin{equation} \label{eq:EE}
\Box h^{\mu\nu} = \frac{16\pi G}{c^4}\,\tau^{\mu\nu} ,
\end{equation}%
with $ \Box \equiv \eta^{\alpha\beta} \partial_{\alpha\beta} $ being the
flat d'Alembertian and $ G $ the Newtonian constant. The pseudo
stress-energy tensor $ \tau^{\mu\nu} $, accounting for the matter as
well as the gravitational fields in harmonic coordinates, is given by
\begin{equation} \label{eq:tau}
\tau^{\mu\nu} = (- g) \,T^{\mu\nu}+\frac{c^4}{16\pi G}\,
\Lambda^{\mu\nu}[h^{\alpha\beta}] ,
\end{equation}%
where $ T^{\mu\nu} $ is the stress-energy tensor of the extended
localized (bounded) matter source with spatially compact support, and $
\Lambda^{\mu\nu} [h^{\alpha\beta}] $ is the gravitational stress-energy
distribution, which depends non-linearly on $ h^{\alpha\beta} $ and its
space-time derivatives. The pseudo-tensor is conserved in the usual
sense:
\begin{equation} \label{eq:divergence_tau}
\partial_\nu\tau^{\mu\nu} = 0 .
\end{equation}%

Now, we are interested in the general expression of the post-Newtonian
expansion valid in the source's near zone, \textit{i.e.} for the
coordinate distance $ r $ significantly smaller than the gravitational
wave length of the emitted radiation. The field
equations~\eqref{eq:harmonicity}--\eqref{eq:EE}, supplemented by a
condition of no-incoming radiation, are solved iteratively as a formal
expansion when $ c \rightarrow + \infty $. The post-Newtonian series of
the field variable $ h^{\mu\nu} $, denoted by means of an overbar, $
\overline{h}^{\mu\nu} = \mathrm{PN}\left[h^{\mu\nu}\right] $, takes the
following ``polylogarithmic'' form \cite{BD86}:
\begin{equation} \label{eq:h_PN_structure}
\overline{h}^{\mu\nu}(\mathbf{x},t,c)=\sum \frac{(\ln
c)^q}{c^p}h^{\mu\nu}_{p,q}(\mathbf{x},t) ,
\end{equation}%
with post-Newtonian coefficients $ h^{\mu\nu}_{p,q} $ ($ p, q \in
\mathbb{N} $) depending on the field point $ \mathbf{x} $ and coordinate
time $ t $. The pseudo-tensor, $ \tau^{\mu\nu} $, admits a post-Newtonian
expansion $ \overline{\tau}^{\mu\nu} $ of the same type, say with
coefficients $ \tau^{\mu\nu}_{p,q} $. The post-Newtonian expansions $
\overline{h}^{\mu\nu} $ and $ \overline{\tau}^{\mu\nu} $ are connected
\textit{via} the field equations by
\begin{equation} \label{eq:EE_PN}
\Box \overline{h}^{\mu\nu} = \frac{16\pi G}{c^4}\,\overline{\tau}^{\mu\nu} ,
\end{equation}%
completed with the harmonic gauge condition
\begin{equation} \label{eq:gauge}
\partial_\nu \overline{h}^{\mu\nu}=0 .
\end{equation}%

In the present paper, we shall not consider expressions of the
post-Newtonian coefficients at a given order, but rather work with whole
post-Newtonian expansions $ \overline{h}^{\mu\nu} $ and $
\overline{\tau}^{\mu\nu} $ of the type \eqref{eq:h_PN_structure}. Such
expansion series will be manipulated by means of the rules of formal
series. In particular, we shall not investigate the eventual convergence
properties, or, more generally, the mathematical nature of these series.
The developments presented in paper II and the present one will thus
constitute some \textit{formal} generalization, to any post-Newtonian
order, of the expressions of the near zone metric and the radiation
reaction force. These expressions have been explicitly
validated by detailed computations at 2PN order for the matching process
as well as the multipole moments \cite{B95}, and at 1.5PN relative order for
the radiation reaction contributions \cite{B93,B97}.

The strategy adopted in paper II was (i) to solve the wave equation
\eqref{eq:EE_PN} in the form of a certain functional of the pseudo tensor, $
\overline{h}^{\mu\nu} \left[\tau^{\alpha\beta}\right] $, and (ii) to verify
that the gauge condition \eqref{eq:gauge} is satisfied as a consequence of the
conservation of $ \tau^{\mu\nu} $, Eq.~\eqref{eq:divergence_tau}. Importantly,
the post-Newtonian expansion of the gravitational field, $
\overline{h}^{\mu\nu} $, was found to be a functional not only of the
post-Newtonian expansion of the pseudo-tensor, $ \overline{\tau}^{\mu\nu} $,
but also of its multipole expansion, denoted below by $
\mathcal{M}(\tau^{\mu\nu}) $, which is valid in the source's exterior, and in
particular in the asymptotic regions at infinity. This can be understood by
the fact that the post-Newtonian solution (comprising the radiation reaction
terms) depends on the boundary conditions imposed at infinity, especially the
no-incoming radiation condition at past null infinity, which must describe a
matter system isolated from the rest of the universe.

In the formalism of paper II, the no-incoming radiation condition is trivially
satisfied, because it is assumed that the matter source is stationary before a
finite date in the remote past, say, for $ t\leq -\mathcal{T} $, with
$ -\mathcal{T} $ denoting a fixed instant in the past. More precisely, we
suppose that both $ h^{\mu\nu} $ and $ \tau^{\mu\nu} $ are \textit{constant}
(with respect to time) outside the causal future development of the source
when it starts emitting radiation, \textit{i.e.} for instance
\begin{equation} \label{eq:paststat}
\tau^{\mu\nu}(\mathbf{x},t) =
\tau_\text{s}^{\mu\nu}(\mathbf{x})~~\text{when}~~t-\frac{\vert\mathbf{x}
\vert}{c}\leq -\mathcal{T} ,
\end{equation}%
where the function $ \tau_\text{s}^{\mu\nu}(\mathbf{x}) $ does not
depend on $ t $, and refers to the ``stationary'' value taken by the
pseudo stress-energy tensor at early times. In the present paper, we
shall continue to assume that Eq.~\eqref{eq:paststat} and the same
equation for $h^{\mu \nu}$ hold. It will be convenient to decompose the
pseudo tensor into the sum of its stationary value, defined by
Eq.~\eqref{eq:paststat}, and a ``non-stationary'' part, denoted by $
\tau^{\mu\nu}_\text{ns}(\mathbf{x} ,t) $, that \textit{vanishes} when $
t\leq -\mathcal{T} $. Hence, we pose, at any time $t$,
\begin{equation} \label{eq:pastns}
\tau^{\mu\nu}(\mathbf{x},t) = \tau_\text{s}^{\mu\nu}(\mathbf{x}) +
\tau_\text{ns}^{\mu\nu}(\mathbf{x},t) .
\end{equation}%
Such splitting is always possible thanks to the hypothesis of past
stationarity. In this formalism, the
derivation of any physical result is to be done under the assumption
\eqref{eq:paststat}, but at the end of the calculation, once the
result is in hand, $\mathcal{T}$ can be sent to $ + \infty $.

The main outcome of paper II is that the post-Newtonian solution of the
wave equation \eqref{eq:EE_PN}, and satisfying \eqref{eq:gauge},
together with correct boundary conditions at infinity, may be written as
the sum of a \textit{particular} (inhomogeneous) solution plus a
specific \textit{homogeneous} solution; both of which are now described.

Let us first consider the particular solution. It is given by the so-called
operator of \textit{instantaneous} potentials, denoted by $ \mathcal{I}^{-1}
$, which is defined by a straightforward iteration of the inverse Poisson
operator $ \Delta^{-1} $. The instantaneous operator $ \mathcal{I}^{-1} $ is
endowed with a regularization, dealing with the boundary of the integrals at
infinity, \textit{i.e.} for $ \vert \mathbf{x}' \vert \rightarrow + \infty $.
This regularization is defined from a particular \textit{finite part}
procedure, indicated by an operator $ \mathop{\mathrm{FP}}_{B = 0} $, with
analytic continuation in the parameter $ B \in \mathbb{C} $ down to $ B=0 $.
Essentially, it consists in multiplying the integrand of the Poisson integral
by a factor $ \vert \mathbf{x}' \vert^B $, and taking the finite part of the
Laurent expansion of the (analytic continuation of the) integral when $
B\rightarrow 0 $. The analytic continuation is always possible at a given
post-Newtonian order for the particular class of functions considered in paper
II as well as in the present paper (\emph{cf.} Appendix B in paper II). Thanks
to this particular regularization,\,\footnote{Note that although it is
sometimes convenient to call it as such, the scheme is strictly speaking not a
regularization since there is no assumption that some infinite quantity should
be formally discarded.} we can select a solution of the iterated Poisson
equations, that is different from the standard Poisson-type integral and
avoids the appearance of divergences at the boundary at infinity (see paper II
for the details). The crucial point is that the latter construction \emph{is}
a particular solution of these equations. We thus have
\begin{equation} \label{eq:particular_solution}
\widetilde{\mathcal{I}^{-1}}\left[ \,\overline{\tau}^{\mu\nu}\right] =
\sum_{k=0}^{+\infty}\left(\frac{\partial}{c\,\partial
t}\right)^{2k}\widetilde{\Delta^{-k-1}}\left[
\,\overline{\tau}^{\mu\nu}\right],
\end{equation}%
where the $ k $-th iterated Poisson integral reads
\begin{equation} \label{eq:iterated_Poisson}
\widetilde{\Delta^{-k-1}}\left[\,\overline{\tau}^{\mu\nu}\right]
(\mathbf{x},t) = \widetilde{\int}
\frac{d^3\mathbf{x}'}{-4 \pi} \,\frac{\vert\mathbf{x}
-\mathbf{x}'\vert^{2k-1}}{(2k)!}\,\overline{\tau}^{\mu\nu} (\mathbf{x}',t) .
\end{equation}%
The overtilde on a quantity means here (following the notation of paper
II) regularization by use of the finite part operation $
\mathop{\mathrm{FP}}_{B = 0} $:
\begin{subequations}\label{overtilde}
\begin{align}
\widetilde{\Delta^{-k-1}}\left[f(\mathbf{x})\right] &\equiv
\mathop{\mathrm{FP}}_{B = 0} \Delta^{-k-1}\left[
\,\vert\mathbf{x}/r_{0}\vert^B\,f(\mathbf{x})\right] , \\ \widetilde{\int}
d^3\mathbf{x}'\,g(\mathbf{x}') &\equiv \mathop{\mathrm{FP}}_{B = 0} \int
d^3\mathbf{x}'\,\vert\mathbf{x}'/r_{0}\vert^B\,g(\mathbf{x}') .
\end{align}
\end{subequations}%
Following earlier articles (Refs. \cite{B95,B98mult} and paper II), we
introduce an independent length scale $ r_0 $ inside the regularization
factor in order to adimensionalize it. Anyway, we know that the constant
$r_0$ disappears from the physical metric as soon as the metric is
functionally related to the local matter distribution. As shown in paper
II, Eq.~\eqref{eq:particular_solution} indeed constitutes a solution of
the d'Alembertian equation \eqref{eq:EE_PN}:
\begin{equation}\label{eq:soldalemb}
\Box\left\{ \widetilde{\mathcal{I}^{-1}}\left[
\,\overline{\tau}^{\mu\nu}\right]\right\}=\overline{\tau}^{\mu\nu} .
\end{equation}%

We next add a homogeneous solution in such a way that the post-Newtonian
metric matches the external one, which is globally defined outside the source
in the form of a multipolar expansion. Clearly, such a homogeneous solution
must be regular inside the matter source, and in particular for $ r
\rightarrow 0 $, assuming that $ r=0 $ corresponds to the spatial origin with
respect to which the multipolar expansion is defined. Therefore, the
homogeneous solution must be made up of a sum of retarded (multipolar) waves
\textit{minus} the corresponding advanced waves; it is well known that such
\textit{anti-symmetric} waves are regular in a neighborhood of $ r=0 $, and are
associated with radiation reaction effects.\,\footnote{Here, the qualifier
``anti-symmetric'' refers to the formal structure of the waves, \textit{i.e.}
retarded minus advanced, but not to their real behavior under a
time reversal, which can be very complicated in high post-Newtonian
approximations.} In the Fourier representation, their contribution takes the
form of a superposition of plane monochromatic waves (see below). However,
following paper II, we prefer a parametrization based on certain
multipole-moment functions in the physical domain.

In paper II, we showed that, when the particular solution is taken to be that
of the instantaneous operator, defined above by
Eq.~\eqref{eq:particular_solution}, the multipole-moment functions
parametrizing the homogeneous solution, say $ \mathcal{A}_L^{\mu\nu}(t) $,
split in fact into a sum of two terms,\,\footnote{Here, $ L = i_1\cdots i_\ell
$ stands for a multi-index composed of $ \ell $ spatial indices ($ \ell $
being often referred to as the multipolar order). For the second term in
Eq.~\eqref{eq:A_L}, we employ a slightly different notation than in paper II,
where it is given by $ \mathcal{R}_L^{\mu\nu}(\text{paper II}) = - 4 \pi
\,\R_L^{\mu\nu} $.}
\begin{equation}\label{eq:A_L}
\A^{\mu\nu}_L(t) = \F^{\mu\nu}_L(t) - 4 \pi \,\R^{\mu\nu}_L(t) .
\end{equation}%
The first of these functions, $ \F^{\mu\nu}_L(t) $, turns out to be
equal to the actual \textit{source multipole moment} of the isolated
system, as seen from the exterior region \cite{B98mult}. It is given by
\begin{equation} \label{eq:F_L}
\F^{\mu\nu}_L(t) =
\widetilde{\int}d^3\mathbf{x}'\,\hat{x}'_L\int_{-1}^{+1}dz\,
\delta_\ell(z)\,\overline{\tau}^{\mu\nu}\,\left(\mathbf{x}',t
-z\vert\mathbf{x}'\vert/c\right) ,
\end{equation}%
with $ \hat{x}'_L = \mathrm{STF} \left( x'_{i_1} \cdots x'_{i_\ell}
\right)$ denoting the symmetric-trace-free (STF) part of the product of
$ \ell $ spatial vectors; the function $ \delta_\ell(z) $, whose
integral is normalized to unity, is defined by
\begin{subequations} \label{eq:delta_l}
\begin{align}
& \delta_\ell (z) = \frac{(2\ell+1)!!}{2^{\ell+1} \ell!} \,(1-z^2)^\ell
, \\ & \int_{-1}^{+1}dz\,\delta_\ell (z) = 1 .
\label{eq:normalization_delta}
\end{align}
\end{subequations}%
We emphasize that expression \eqref{eq:F_L}, together with \eqref{eq:delta_l},
is physically valid only for post-Newtonian sources. As such, it must be
considered in a perturbative sense and given as some formal
post-Newtonian expansion. The explicit post-Newtonian series reads (see paper
II)
\begin{equation} \label{eq:F_L_PN}
\F^{\mu\nu}_L(t) = \sum_{j=0}^{+\infty} \left(\frac{d}{c
\,dt}\right)^{\!\!2j}\,\widetilde{\int}
d^3\mathbf{x}'\,\widetilde{\Delta^{-j}}\left(\hat{x}'_L\right)
\,\overline{\tau}^{\mu\nu}\left(\mathbf{x}',t\right) ,
\end{equation}%
where we have introduced the convenient notation
\begin{equation} \label{eq:Delta-k}
\widetilde{\Delta^{-j}}\left(\hat{x}'_L\right) = \frac{(2\ell+1)!!}{2^j j!
(2\ell+2j+1)!!}\,\vert\mathbf{x}'\vert^{2j}\hat{x}'_L ,
\end{equation}%
which is justified by the fact that $ \Delta^j \bigl[
\widetilde{\Delta^{-j}} \left( \hat{x}'_L \right) \bigr] = \hat{x}'_L $.

As can be seen from Eqs.~\eqref{eq:F_L} or \eqref{eq:F_L_PN}, the function $
\F^{\mu\nu}_L $ is defined as a functional of the \textit{post-Newtonian}
expansion of the pseudo-tensor $ \overline{\tau}^{\mu\nu} $, valid in the
source's near zone. The second piece of the multipolar function
\eqref{eq:A_L}, namely $ \R^{\mu\nu}_L $, directly issues from the matching
process to the external field, and arises purely from non-linearities. Thus,
it depends on the expression of the pseudo-tensor in the exterior of the
source (including the region located at infinity), where it is given in the
form of a \textit{multipolar} series $ \mathcal{M}(\tau^{\mu\nu}) $, with $
\mathcal{M} $ representing the multipolar expansion. Actually, in the present
formalism, $ \mathcal{M}(h^{\mu\nu}) $, and $ \mathcal{M}(\tau^{\mu\nu}) $ as
well, are \textit{identified} with the so-called multipolar-post-Minkowskian
expansion of the external field, in the sense of
Ref.~\cite{BD86}.\,\footnote{The post-\textit{Minkowskian} expansion (or
non-linearity expansion when $ G\rightarrow 0 $) plays a crucial role in the
construction of the multipolar-post-Minkowskian solution in Ref.~\cite{BD86},
but shall not be considered for the present purpose. Indeed, we shall assume
here that we are always entitled to sum up formally the post-Minkowskian
series and to treat the formal limit as if it were ``exact''.} The function $
\R_L^{\mu\nu} $ may be written in a compact way as (see \cite{B93} and paper
II)
\begin{equation} \label{eq:R_L}
\R^{\mu\nu}_L(t) = \widetilde{\int} \frac{d^3\mathbf{x}'}{-4\pi}
\,\hat{x}'_L \int_{1}^{+\infty}dz\,
\gamma_\ell(z)\,\mathcal{M}(\tau_\text{ns}^{\mu\nu})\!\left(\mathbf{x}',t
-z \vert\mathbf{x}'\vert/c\right) ,
\end{equation}%
where $\gamma_\ell(z) \equiv -2\,\delta_\ell(z)$ is such that
\begin{equation}\label{eq:gamma_l}
\int_{1}^{+\infty}dz\, \gamma_\ell (z) = 1 .
\end{equation}%
Remark that the above normalization of the integral of $ \gamma_\ell(z)
$ is a consequence of that of $ \delta_\ell(z) $
[Eq.~\eqref{eq:normalization_delta}], together with the fact that
$ \int_{-\infty}^{+\infty}dz \, \gamma_\ell (z) = 0 $ by analytic
continuation in the variable $ \ell\in\mathbb{C} $.

Note that, as indicated in Eq.~\eqref{eq:R_L}, the function $ \R_L^{\mu\nu} $
involves the \textit{non-stationary} part of the pseudo-tensor, $
\tau_\text{ns}^{\mu\nu} $, given by
Eqs.~\eqref{eq:paststat}--\eqref{eq:pastns}; hence the integral over $ z $ is
well-defined, as it ranges in fact on a finite domain $ \left[ 1,
z_\mathrm{max} \right] $, where $ z_\mathrm{max} + 1 = c\, (t+\mathcal{T})/
\vert \mathbf{x}' \vert $. This means that $ \R_L^{\mu\nu} $ depends on the
initial instant of emission, $ -\mathcal{T} $, but since it appears only
through an anti-symmetric type wave, in Eq.~\eqref{eq:h_PN_new} below, the
dependence on $ -\mathcal{T} $ finally cancels out.

One can show that an alternative form for $ \R^{\mu\nu}_L $ is
\begin{equation} \label{eq:R_L_explicit}
\R^{\mu\nu}_L(t) = - c^{2\ell + 1} (2\ell+1)!! \widetilde{\int}
\frac{d^3\mathbf{x}'}{-4\pi}\,\hat{\partial}'_L \!\left(
\frac{1}{\vert\mathbf{x}'\vert}\mathop{\mathcal{M}
(\tau_\text{ns}^{\mu\nu})}^{(-2\ell-1)}(t-
\vert\mathbf{x}'\vert/c)\right) ,
\end{equation}%
where $ \hat{\partial}'_L $ is the STF multi-derivative operator
associated with the integration point $ \mathbf{x}' $, and the superscript
$ (-2 \ell - 1) $ over $ \mathcal{M} (\tau_\text{ns}^{\mu\nu}) $
denotes the $ (2 \ell + 1) $-th time \textit{anti-derivative} that
vanishes in the remote past, say, when $ t\leq -\mathcal{T} $.
The STF spatial multi-derivative admits the following explicit
expression:
\begin{equation} \label{eq:monopole_derivatives}
\hat{\partial}_L \!\left( \frac{f\left(t-r/c\right)}{r} \right) =
(-1)^\ell\,\hat{n}_L\!\sum_{i=0}^{\ell} \frac{(\ell+i)!}{2^i i!
(\ell-i)!}\frac{f^{(\ell-i)}(t-r/c)}{c^{\ell-i}\,r^{i+1}} ,
\end{equation}%
where $ r=\vert\mathbf{x}\vert $, $ \hat{n}_L = \hat{x}_L/r^\ell $, $
\hat{\partial}_L = \mathrm{STF} \left(\partial_{i_1} \cdots
\partial_{i_\ell}\right) $; the superscript $ (\ell - i) $ means the
$(\ell -i )$-th time derivative.

With $ \A^{\mu\nu}_L $ being the sum of these functions as given by
Eq.~\eqref{eq:A_L}, the post-Newtonian expansion in the form obtained
using the method of matched asymptotic expansions in paper II reads:
\begin{equation} \label{eq:h_PN}
\overline{h}^{\mu\nu}=\frac{16\pi
G}{c^4}\widetilde{\mathcal{I}^{-1}}\left[\,
\overline{\tau}^{\mu\nu}\right] - \frac{4G}{c^4}
\sum^{+\infty}_{\ell=0} \frac{(-1)^\ell}{\ell!} \,\hat{\partial}_L
\!\left\{ \frac{\A^{\mu\nu}_L (t-r/c)-\A^{\mu\nu}_L(t+r/c)}{2\,r}
\right\} . 
\end{equation}%
The explicit expression of the multipolar waves can be written from
Eq.~\eqref{eq:monopole_derivatives} (with $ c \rightarrow - c $ in the case of
the advanced wave), but we shall give some alternative formulas in
Eqs.~\eqref{eq:antisymmetric_F_wave} and \eqref{eq:Fourier} below.

\section{New expression of the post-Newtonian expansion}
\label{sec:new_PN_expansion}

The idea for determining the new form of the post-Newtonian expansion is based
on the remark that, in Eq.~\eqref{eq:h_PN}, the part of the multipolar
anti-symmetric wave that is associated with the function $ \F_L^{\mu\nu} $
corresponds in fact to the ``odd'' parity part (\textit{i.e.}, the part
composed of odd powers of $1/c$) of the retardation expansion in the standard
retarded integral. Since the operator of the instantaneous potentials
\eqref{eq:particular_solution}--\eqref{eq:iterated_Poisson} precisely equates
the corresponding ``even'' part, we can reconstitute the complete retarded
integral of the post-Newtonian source term in a nice way.

To begin with, we notice that the post-Newtonian expansion of the
multipolar anti-symmetric wave reads
\begin{equation} \label{eq:antisymmetric_F_wave}
\hat{\partial}_L \!\left\{\frac{\F^{\mu\nu}_L
(t-r/c)-\F^{\mu\nu}_L(t+r/c)}{2r} \right\}
= -\frac{1}{(2\ell+1)!!}\,\sum_{p=0}^{+\infty}
\widetilde{\Delta^{-p}}\,\left(\hat{x}_L\right)\,
\left(\frac{d}{c\,dt}\right)^{2p+2\ell+1} \!\!\! 
\F^{\mu\nu}_L(t) , 
\end{equation}%
with notation \eqref{eq:Delta-k}. Next, we recall that the function
$ \F^{\mu\nu}_L(t) $ itself is given by the expression \eqref{eq:F_L} or
alternatively by the post-Newtonian series \eqref{eq:F_L_PN}. After
inserting Eq.~\eqref{eq:F_L_PN} into \eqref{eq:antisymmetric_F_wave}, we
are led to a rather complicated double expansion. Furthermore, we have
to sum up the latter result over all multipolarities $ \ell=0, \cdots, +
\infty $. At this point, let us change in an appropriate way the
summation indices (posing in particular $ k=j+p+\ell $). We are then
able to recognize a double summation which turns out to be nothing but
the expression of the quantity $ \vert \mathbf{x} - \mathbf{x}'
\vert^{2k} $, when developed in terms of STF harmonics. It can be
written in a particularly economic form according to the following
formula, in which we resort again to notation \eqref{eq:Delta-k}:
\begin{equation} \label{eq:multipole_r2k}
\frac{\vert\mathbf{x}-\mathbf{x}'\vert^{2k}}{(2k+1)!}
=\sum_{\ell=0}^{k}\frac{(-1)^\ell}{\ell!\,(2\ell+1)!!}
\sum_{p=0}^{k-\ell}\widetilde{\Delta^{-p}}\left(\hat{x}_L\right)
\widetilde{\Delta^{-k+\ell+p}}\left(\hat{x}'_L\right) .
\end{equation}%
This formula can be viewed as a finite expansion in terms of Legendre
polynomials, since the product $ \widetilde{\Delta^{-p}}
\left(\hat{x}_L\right) \widetilde{\Delta^{-q}} \left(\hat{x}'_L\right) $
(where $ p, q \in \mathbb{N} $) is proportional to $ \hat{x}_L \,
\hat{x}'_L $, and the latter is known to be in turn proportional to the
$ \ell $-th Legendre polynomial of the scalar products of the unit
vectors $\mathbf{n} = \mathbf{x}/r $ and $ \mathbf{n}' = \mathbf{x}'/r'
$, namely $ P_\ell \left( \mathbf{n} \cdot \mathbf{n}' \right) $. Thanks
to Eq.~\eqref{eq:multipole_r2k}, we arrive at the interesting
relation
\begin{equation} \label{eq:relation_F_wave_iterated_Poisson}
\sum^{+\infty}_{\ell=0} \frac{(-1)^\ell}{\ell!} \,\hat{\partial}_L
\!\left\{\frac{\F^{\mu\nu}_L (t-r/c)-\F^{\mu\nu}_L(t+r/c)}{2\,r}
\right\} = - \sum_{k=0}^{+\infty}\left(\frac{\partial}{c\,\partial
t}\right)^{\!2k+1}\,\widetilde{\int}
d^3\mathbf{x}'\,\frac{\vert\mathbf{x} -\mathbf{x}'\vert^{2k}}{(2k+1)!}\,
\overline{\tau}^{\mu\nu} (\mathbf{x}',t) ,
\end{equation}%
which shows, as announced before, that the anti-symmetric wave
associated with $ \F^{\mu\nu}_L(t) $ is equal to the odd part
[\textit{i.e.} corresponding to $ n = 2k + 1 $, $ k \in \mathbb{N} $, in
Eq.~\eqref{eq:box-1_tilde} below] of the expansion of the retarded
integral as regularized by means of the finite part FP.

In order to gather terms \eqref{eq:particular_solution} and
\eqref{eq:relation_F_wave_iterated_Poisson}, which are both present in
Eq.~\eqref{eq:h_PN}, we need to consider the new object that is made of
the formal expansion of the standard retarded ($ \mathrm{R} $) integral
$ \Box_\mathrm{R}^{-1} $, but where each of the terms is regularized by
means of the FP procedure (indicated by an overtilde) to deal with the
divergences appearing on the boundary at infinity:
\begin{equation} \label{eq:box-1_tilde}
\widetilde{\Box_\mathrm{R}^{-1}}\left[\,\overline{
\tau}^{\mu\nu}\right](\mathbf{x},t) \equiv
\sum_{n=0}^{+\infty}\frac{(-1)^n}{n!}\left(\frac{\partial}{c\,\partial
t}\right)^{\!n}\,\widetilde{\int}
\frac{d^3\mathbf{x}'}{-4\pi}\,\vert\mathbf{x}
-\mathbf{x}'\vert^{n-1}\,\overline{\tau}^{\mu\nu}(\mathbf{x}',t) .
\end{equation}%
This object should be carefully distinguished from, say, $
\Box_\mathrm{R}^{-1} \left[\, \overline{\tau}^{\mu\nu} \right] $,
involving the retarded integral $ \Box_\mathrm{R}^{-1} $ without
any kind of regularization and with the retardations being kept
unexpanded inside. Eq.~\eqref{eq:box-1_tilde} is also \textit{a priori}
different from the more ``global'' solution $ \Box_\mathrm{R}^{-1}
\left[ \tau^{\mu\nu} \right] = c^4 \,h^{\mu\nu}/(16 \pi G) $, in which
the pseudo-tensor has not been expanded in post-Newtonian fashion.
As we shall see below, Eq.~\eqref{eq:box-1_tilde} differs also from the
post-Newtonian solution itself, \textit{i.e.} $
\widetilde{\Box_\mathrm{R}^{-1}} \left[ \overline{\tau}^{\mu\nu} \right]
\not= c^4 \,\overline{h}^{\mu\nu}/(16 \pi G) $. 

Let us emphasize that Eq.~\eqref{eq:box-1_tilde} constitutes merely the
\textit{definition} of a formal Taylor-type post-Newtonian expansion, each
term of which is built from the (also formal) post-Newtonian expansion of the
pseudo-tensor. Such a definition is of interest as it corresponds to what one
would intuitively think is the ``natural'' way of performing the
post-Newtonian iteration, \textit{i.e.} by Taylor expanding the retardations.
Moreover, each of the terms of the series \eqref{eq:box-1_tilde} are
mathematically well defined thanks to the FP process, and can therefore be
implemented in practical computations like in the one reported in
Ref.~\cite{NB05}. The important point, and the only one which matters for our
purpose, is that the definition \eqref{eq:box-1_tilde} solves, in a
post-Newtonian sense, the required wave equation, namely
\begin{equation}
\Box\left\{ \widetilde{\Box_\mathrm{R}^{-1}}\left[
\,\overline{\tau}^{\mu\nu}\right]\right\}=\overline{\tau}^{\mu\nu} ,
\end{equation}%
and constitutes therefore a legitimate particular solution for
determining the general form of the post-Newtonian expansion $
\overline{h}^{\mu\nu} $ as the sum of this quantity and
some homogeneous solution which is necessarily of the anti-symmetric
type.

Combining Eq.~\eqref{eq:relation_F_wave_iterated_Poisson} with our
definition of the operator of the instantaneous potentials
\eqref{eq:particular_solution}-\eqref{eq:iterated_Poisson}, which
corresponds to the even part of the retardation expansion [\textit{i.e.}
$ n = 2k $, $ k \in \mathbb{N} $, in Eq.~\eqref{eq:box-1_tilde}], we find
\begin{equation} \label{eq:box_I}
\widetilde{\Box_\mathrm{R}^{-1}}\left[ \overline{\tau}^{\mu\nu}\right] =
\widetilde{\mathcal{I}^{-1}}\left[ \,\overline{\tau}^{\mu\nu}\right]
-\frac{1}{4\pi} \sum^{+\infty}_{\ell=0} \frac{(-1)^\ell}{\ell!}
\hat{\partial}_L \left\{ \frac{\F^{\mu\nu}_L
(t-r/c)-\F^{\mu\nu}_L(t+r/c)}{2\,r} \right\} . 
\end{equation}%
This formula is the basis of our writing of the new form of the
post-Newtonian expansion. Indeed, by comparing Eqs.~\eqref{eq:h_PN} and
\eqref{eq:box_I}, we obtain the following expression:
\begin{equation}
\label{eq:h_PN_new}
\overline{h}^{\mu\nu} = \frac{16\pi G}{c^4} \Biggl[
\widetilde{\Box_\mathrm{R}^{-1}} \left[ \overline{\tau}^{\mu\nu}\right]
+ \sum^{+\infty}_{\ell=0} \frac{(-1)^\ell}{\ell!} \hat{\partial}_L
\left\{ \frac{\R^{\mu\nu}_L (t-r/c)-\R^{\mu\nu}_L(t+r/c)}{2\,r} \right\}
\Biggr] ,
\end{equation}%
which relates the post-Newtonian expansion of $ h^{\mu\nu} = (16 \pi G/c^4)
\,\Box_\mathrm{R}^{-1} \left[ \tau^{\mu\nu} \right] $ to the developed
operator $ \widetilde{ \Box_\mathrm{R}^{-1}} [ \overline{\tau}^{\mu\nu} ] $
acting on the post-Newtonian expansion of the source term, $
\overline{\tau}^{\mu\nu} $. Eq.~\eqref{eq:h_PN_new}, together with the
conservation law for the pseudo-tensor [Eq.~\eqref{eq:divergence_tau}],
constitute our final solution for the present problem.

Let us insist that the formula \eqref{eq:h_PN_new} should be understood, for
practical calculations, in an iterative post-Newtonian way. At the lowest
order, the source has compact support and there is no need of the finite part
procedure. Higher-order equations are obtained by inserting previous
iterations into the source term $ \overline{\tau}^{\mu\nu} $ [or $ \mathcal{M}
(\tau^{\mu\nu}) $] truncated at this order; the source will be of non-compact
support because of the gravitational field contribution. One then multiplies
it by the regularization factor $ \vert\mathbf{x}/r_0\vert^B $,
integrates and performs the Laurent expansion when $ B \to 0 $ to get the
finite parts needed in the two terms of Eq.~\eqref{eq:h_PN_new}. Some examples
of such finite parts of integrals can be found in paper I [see \textit{e.g.}
Eqs.~(4.15--16) there].

Equation~\eqref{eq:h_PN_new} is more interesting and probably more
``fruitful'' in applications than the old form \eqref{eq:h_PN}. Indeed, recall
that the anti-symmetric waves in Eq.~\eqref{eq:h_PN}, parametrized by $
\A^{\mu\nu}_L = \F^{\mu\nu}_L - 4 \pi \R^{\mu\nu}_L $, are linked with
radiation reaction effects. It was noticed in paper II that the part associated
with $ \F^{\mu\nu}_L $ starts at the usual dominant order for the radiation
reaction force, \textit{i.e.} the 2.5PN order (where it yields the standard
expression for the ``Newtonian'', quadrupolar radiation reaction). It includes
also the next term of order 3.5PN corresponding to the mass octupole and
current quadrupole radiation reaction, and probably also the subsequent
contributions of higher multipole moments that parametrize the radiation
reaction at the linearized level.\,\footnote{By multipole moments here, we
mean the ``\textit{source}'' moments defined in Ref.~\cite{B98mult}, which
couple together non-linearly in the expressions of the radiation reaction
force and the ``\textit{radiative}'' moments seen at infinity.} On the other
hand, the part linked with the functions $ \R^{\mu\nu}_L $ has been shown in
Ref.~\cite{B93} to be due to the \textit{non-linear} contribution of the
gravitational wave tails in the radiation reaction force. It was determined
beforehand in Ref.~\cite{BD88} and shown there to appear at the 4PN order,
\textit{i.e.} 1.5PN order relative to the dominant radiation reaction effect.

Consequently, when computing the radiation reaction force limited at the
3.5PN order, the part associated with $ \R^{\mu\nu}_L $ can be
neglected, and we can keep only the part given by the first term of
Eq.~\eqref{eq:h_PN_new}; this corresponds indeed to the intuitive way of
performing the post-Newtonian iteration (bypassing the tail effects),
which was advocated in old investigations of the post-Newtonian
expansion for isolated fluids, like those of Ref.~\cite{ADec75}. Thus,
Eq.~\eqref{eq:h_PN_new} rigorously proves that the 3.5PN radiation
reaction must be computed by considering the odd-parity part of the
retardations in the usual retarded integral, provided that the source
terms be multiplied by the regularization factor $
\vert\mathbf{x}'\vert^B $ and the integral computed following the FP
prescription when $ B\rightarrow 0 $. This is exactly what has been done
in the paper \cite{NB05} (see notably Section IV there) devoted to the
computation of the 3.5PN radiation reaction in the case of compact
binary systems. However, for higher-order computations, at the 4PN level
and beyond, it is definitely necessary to take into account the
contribution of the functions $ \R^{\mu\nu}_L $ in
Eq.~\eqref{eq:h_PN_new} which is induced by wave tails.

For some applications, it may be useful to rephrase our result with the help
of some more systematic notation. Notably, we may use an appropriate operator,
denoted $ \mathcal{D}^{-1} $ henceforth, which maps the source term of the
Einstein field equations, namely the pseudo tensor $ \tau^{\mu\nu} $, onto the
post-Newtonian solution given by Eq.~\eqref{eq:h_PN_new}. Let us first recall
that $ h^{\mu\nu} = (16 \pi G/c^{4})\, \Box_\mathrm{R}^{-1} \left[
\tau^{\mu\nu} \right] $, which allows us to suppress the overall factor $16 \pi
G/c^{4}$ in Eq.~\eqref{eq:h_PN_new} and reexpress the latter equation in the
form of
\begin{equation}\label{eq:res_new}
\overline{\Box_\mathrm{R}^{-1} \left[\tau^{\mu\nu}\right]} =
\widetilde{\Box_\mathrm{R}^{-1}} \left[ \overline{\tau}^{\mu\nu}\right]
+ \Hom^{\mu\nu} ,
\end{equation}%
where the specific homogeneous, anti-symmetric wave type solution
is defined by
\begin{equation} \label{eq:H}
\Hom^{\mu\nu} \equiv \sum_{\ell = 0}^{+ \infty}
\frac{(-1)^\ell}{\ell!} \hat{\partial}_L \left\{
\frac{\R_L^{\mu\nu}(t-r/c) - \R_L^{\mu\nu}(t+r/c)}{2r} \right\} .
\end{equation}%
Eq.~\eqref{eq:res_new} suggests introducing some convenient
\textit{functional} operators acting on the components of $ \tau^{\mu\nu} $.
For the first term of Eq.~\eqref{eq:res_new}, we introduce the operator $
\overline{\Box_\mathrm{R}^{-1}} $, whose action on $ \tau^{\mu\nu} $ is
defined by
\begin{equation}\label{eq:res_new1}
\overline{\Box_\mathrm{R}^{-1}} [\tau^{\mu\nu}] \equiv
\widetilde{\Box_\mathrm{R}^{-1}} \left[\overline{\tau}^{\mu\nu}\right] .
\end{equation}%
With this notation, we include into the definition of $
\overline{\Box_\mathrm{R}^{-1}} $ both the FP process, indicated by the
overtilde [see Eq.~\eqref{overtilde}], and the operation of taking the
post-Newtonian expansion, $
\overline{\tau}^{\mu\nu}=\mathrm{PN}[\tau^{\mu\nu}] $. For the second term of
Eq.~\eqref{eq:res_new}, we pose $ \Hom[\tau^{\mu\nu}] \equiv \Hom^{\mu\nu} $,
and hence $\Hom$ represents the particular functional that is obtained when
the functions $ \R_L^{\mu\nu} $ in Eq.~\eqref{eq:H} are replaced by their
explicit expressions in terms of $ \tau^{\mu\nu} $ as given by
Eqs.~\eqref{eq:R_L} or \eqref{eq:R_L_explicit}. Note that $ \R_L^{\mu\nu} $
and $\Hom^{\mu\nu} $ depend in fact on the multipole expansion
$\mathcal{M}(\tau^{\mu\nu}) $ of the pseudo tensor, rather than on $
\tau^{\mu\nu} $ itself. As a result, the operation of taking the multipole
expansion ($ \mathcal{M} $) is included into the above definition of the
functional $ \Hom[\tau^{\mu\nu}] $. Naturally, the latter functional is also a
function of the field point $ \mathbf{x} $ and of (coordinate) time $ t $.
Thus, our notation means more precisely $ \Hom[\tau^{\mu\nu}](\mathbf{x},t)
\equiv \Hom^{\mu\nu}(\mathbf{x},t) $. In a similar way, it is convenient to
make use of what we call $ \R_L$, namely a functional of $ \tau^{\mu\nu} $ and
a function of time $ t $ defined by $ \R_L[\tau^{\mu\nu}](t) \equiv
\R_L^{\mu\nu}(t) $. We can then rewrite Eq.~\eqref{eq:res_new} as
\begin{equation}\label{eq:res_D}
\overline{\Box_\mathrm{R}^{-1} \left[\tau^{\mu\nu}\right]} =
\mathcal{D}^{-1} \left[ \tau^{\mu\nu}\right] ,
\end{equation}%
where the operator $ \mathcal{D}^{-1} $ is naturally split into the two
terms appearing in Eq.~\eqref{eq:res_new} as
\begin{equation}\label{eq:D}
\mathcal{D}^{-1} \equiv \overline{\Box_\mathrm{R}^{-1}} + \Hom .
\end{equation}%
We shall use this operator in the next Section devoted to the
harmonicity conditions.

Note that the anti-symmetric multipolar waves such as $ \Hom^{\mu\nu} $,
given in Eq.~\eqref{eq:H}, can be represented as a Fourier superposition of
plane waves. We give here the relevant formulas without proof. The Fourier
representation is arguably simpler and perhaps more physically appealing, but
the alternative STF multipolar anti-symmetric formulation \eqref{eq:H}, which
we mostly use in the present work, displays a structure that is often more
useful in applications. Both forms are linked by
\begin{equation}
\label{eq:Fourier}
\Hom^{\mu\nu}(\mathbf{x}, t) = \int d^3\mathbf{k}
\left[A_\text{out}^{\mu\nu}(\mathbf{k}) \, e^{- 2 \pi i k c t} +
A_\text{in}^{\mu\nu}(\mathbf{k}) \, e^{2 \pi i k c t} \right] e^{2 \pi i
\mathbf{k} \cdot \mathbf{x}} ,
\end{equation}%
where the Fourier amplitudes $ A_\text{out}^{\mu\nu}(\mathbf{k}) $ and $
A_\text{in}^{\mu\nu}(\mathbf{k}) $ read, in terms of the multipolar
functions $ \R^{\mu\nu}_L(t) $, as
\begin{subequations}
\begin{align}
A_\text{out}^{\mu\nu}(\mathbf{k}) &= - \frac{c}{2k i} \sum_{\ell =
0}^{+\infty} \frac{(-2\pi i k)_{<L>}}{\ell!} \, F(\R_L^{\mu\nu})(c k) ,
\\ A_\text{in}^{\mu\nu}(\mathbf{k}) &= \frac{c}{2k i} \sum_{\ell =
0}^{+\infty} \frac{(-2\pi i k)_{<L>}}{\ell!} \, F(\R_L^{\mu\nu})(- c k)
,
\end{align}
\end{subequations}%
with $ F(\R_L^{\mu\nu})(\omega) \equiv \int_{-\infty}^{+\infty} dt \,
e^{2 \pi i \omega t} \,\R_L^{\mu\nu}(t) $ representing the
one-dimensional Fourier transform of $ \R_L^{\mu\nu}(t) $. Here, we pose
$ k\equiv\vert\mathbf{k}\vert $, $ k_L\equiv k_{i_1}\cdots k_{i_\ell} $
and the brackets surrounding indices denote the STF projection.
Conversely, the function $ \R^{\mu\nu}_L(t) $ is entirely determined by
means of the Fourier amplitudes as\,\footnote{This relation is easily
checked to be true by virtue of the formula
\begin{equation}
 \left(\frac{\partial}{c \partial t}\right)^{2 \ell + 1} \R^{\mu\nu}_L(t) =
(2\ell + 1 )!! (-1)^{\ell+1} (\hat{\partial}_L \Hom^{\mu\nu})(\mathbf{0}, t) ,
\end{equation}%
where
$ \Hom^{\mu\nu}(\mathbf{x}, t) $ is given by \eqref{eq:H}.}
\begin{equation}
\R_L^{\mu\nu}(t) = (2\ell + 1 )!! \int d^3\mathbf{k} \, \frac{(- 2 \pi i
k)_{<L>}}{(- 2\pi i k)^{2\ell + 1}} \left[-
A_\text{out}^{\mu\nu}(\mathbf{k}) \, e^{- 2 \pi i k c t} +
A_\text{in}^{\mu\nu}(\mathbf{k}) \, e^{2 \pi i k c t} \right] .
\end{equation}%

\section{Harmonicity condition} \label{sec:harmonicity}

In this Section, we prove that our post-Newtonian solution \eqref{eq:h_PN_new}
is divergenceless as a consequence of the harmonicity condition, and thus
constitutes an actual solution of the Einstein field equations. More
precisely, we check that the conservation of the pseudo-tensor,
Eq.~\eqref{eq:divergence_tau}, implies the satisfaction of the harmonicity
conditions~\eqref{eq:gauge}. This proof has already been done in paper II, but
we present here the main steps of an alternative demonstration, using some of
the tools employed there, and based directly on the new form
\eqref{eq:h_PN_new} of the post-Newtonian expansion.

In order to verify the harmonicity condition, it is convenient to establish a
sufficient property involving the operator $ \mathcal{D}^{-1} $ introduced in
Eq.~\eqref{eq:D}. In fact, the solution written in the form \eqref{eq:res_D}
will be divergenceless as a consequence of $ \partial_\nu \tau^{\mu\nu} =0 $,
if (mathematically speaking) we are allowed to commute the partial derivative
$ \partial_\nu $ with the $ \mathcal{D}^{-1} $ operator, \textit{i.e.} $
\partial_\nu \mathcal{D}^{-1} \left[ \tau^{\mu\nu}\right] = \mathcal{D}^{-1}
\left[ \partial_\nu \tau^{\mu\nu}\right]=0 $. Let us define the following
\textit{commutator}: $ \bigl[\partial_\nu, \,\mathcal{D}^{-1}\bigr] \equiv
\partial_\nu\,\mathcal{D}^{-1}-\mathcal{D}^{-1}\,\partial_\nu $. Thus, it is
sufficient to prove that
\begin{equation}\label{eq:commutator}
\bigl[\partial_\nu, \,\mathcal{D}^{-1}\bigr] = 0 .
\end{equation}%
The commutator corresponding to the partial time derivative is seen to be
zero, $ \bigl[\partial/\partial t, \,\mathcal{D}^{-1}\bigr]=0 $, because the
operations of taking the spatial integral and performing spatial
differentiations which are involved in $ \mathcal{D}^{-1} $ are obviously
transparent to the time derivative. All what remains for us to prove is that $
\bigl[\partial_i, \,\mathcal{D}^{-1}\bigr] = 0 $, for the partial
\textit{space} derivative $ \partial_i=\partial/\partial x^i $. Therefore, we
shall now show that
\begin{equation}\label{eq:commutator2}
\bigl[\partial_i, \,\Hom\bigr] = - \bigl[\partial_i,
\,\overline{\Box_\mathrm{R}^{-1}}\bigr] .
\end{equation}%

The first stage of our proof consists in evaluating $ \partial_i
\mathcal{D}^{-1}[\tau^{\mu\nu}] $ by decomposing the multi-derivative
acting on the anti-symmetric wave $ \Hom_L^{\mu\nu} \equiv
(2 r)^{-1}[\R_L^{\mu\nu}(t-r/c) - \R_L^{\mu\nu}(t+r/c)] $ into
symmetric-trace-free (STF) pieces using the technical formula
\begin{equation}
\partial_i \hat{\partial}_L = \hat{\partial}_{iL} + \frac{\ell}{2 \ell
+1} \delta_{i\langle i_\ell} \hat{\partial}_{L-1\rangle} \Delta ,
\end{equation}%
where the brackets $ \langle\cdots\rangle $ denote the STF operation for
the indices they enclose. Each multipole order $ \ell $ in the
right-hand side of Eq.~\eqref{eq:H} is thus given by a sum of two terms.
After changing appropriately the summation indices and bearing in mind
that $ \Hom_L^{\mu\nu} $ satisfies the homogeneous wave equation, so
that $ \Delta \Hom_L^{\mu\nu} = (c^{-1}\partial/\partial t)^2
\Hom_L^{\mu\nu} $, we arrive at
\begin{align} \label{eq:di_H}
&\partial_i \Hom[\tau^{\mu\nu}] = \sum_{\ell = 0}^{+ \infty}
\frac{(-1)^\ell}{\ell!} \,\hat{\partial}_L \Biggl\{ \frac{1}{2r} \Biggl[
\ell \, \delta_{i\langle i_\ell} \R_{L-1\rangle}[\tau^{\mu\nu}](t + r/c)
\nonumber \\ & \qquad \qquad + \frac{1}{2 \ell + 3}
\left(\frac{\partial}{c \partial t}\right)^2 \R_{iL}[\tau^{\mu\nu}](t +
r/c)
- \Bigl( \text{\textit{idem} for the
retarded wave}\Bigr) \Biggr] \Biggr\} .
\end{align}%

In a second stage, we shall work on the quantity $ \Hom[\partial_i
\tau^{\mu\nu}] $ in order to obtain an alternative expression involving
$ \R_L[\tau^{\mu\nu}] $ rather than $ \R_L[\partial_i \tau^{\mu\nu}] $.
The idea is to integrate by parts the integral on $ \mathbf{x}' $
defining $ \R_L[\partial_i \tau^{\mu\nu}] $ as \footnote{Henceforth we
pose $r_0=1$.}
\begin{equation} \label{eq:R_L_di_tau}
\R_L[\partial_i \tau^{\mu\nu}] = \mathop{\mathrm{FP}}_{B=0} \int
\frac{d^3\mathbf{x}'}{-4\pi}\,\vert\mathbf{x}'\vert^B \hat{x}'_L
\int_{1}^{+\infty}dz\, \gamma_\ell(z)\,\mathcal{M}(\partial_i
\tau_\text{ns}^{\mu\nu})\!\left(\mathbf{x}',t
-z\vert\mathbf{x}'\vert/c\right) ,
\end{equation}%
in a way that the source term does not contain any more derivatives. To
begin with, we observe that $ \mathcal{M} (\partial_i
\tau^{\mu\nu}_\text{ns}) $ is equal to a total space derivative, plus a
term that turns out to involve a time derivative:
\begin{equation} \label{eq:total_derivatives}
\mathcal{M}(\partial_i \tau^{\mu\nu}_\text{ns}) (\mathbf{x}',t - z
\vert\mathbf{x}'\vert/c) = \partial_i [
\mathcal{M}(\tau^{\mu\nu}_\text{ns})(\mathbf{x}', t - z
\vert\mathbf{x}'\vert/c)] + z \,n'_i \left(\frac{\partial}{c \partial
t}\right) \mathcal{M}(\tau^{\mu\nu}_\text{ns}) (\mathbf{x}',t - z
\vert\mathbf{x}'\vert/c) .
\end{equation}%
The first contribution may be integrated by parts, which has the effect of
transferring the operator $ \partial_i $ onto the product $
\vert\mathbf{x}'\vert^B \hat{x}'_L $. Its explicit action on the latter
quantity reads
\begin{equation}
\partial_i (\vert\mathbf{x}'\vert^B \hat{x}'_L) = \left[B n'_i \hat{n}'_L + 
(2\ell+1) (n'_i \hat{n}'_L - \hat{n}'_{iL}) \right] 
\vert\mathbf{x}'\vert^{B+\ell-1} .
\end{equation}%
As $ \tau^{\mu\nu}_\text{ns} = 0 $ in the remote past, the surface term
does not contribute. By applying the chain rule, we rewrite the second
contribution in the right-hand side of Eq.~\eqref{eq:total_derivatives}
as $ - (z \, x'_i/\vert\mathbf{x}'\vert^2) (\partial/\partial z)
\mathcal{M}(\tau^{\mu\nu}_\text{ns})(\mathbf{x}', t - z
\vert\mathbf{x}'\vert/c) $. In particular, the contribution to the
corresponding integral on $ z $ may be integrated by parts in turn, and
hence we are led to consider the first derivative $ (d/dz)\bigl( z
\gamma_\ell (z)\bigr) $, which verifies $ (d/dz) \bigl( z
\gamma_\ell(z)\bigr) = (2\ell + 1) \bigl(\gamma_\ell(z) - \gamma_{\ell -
1}(z)\bigr) $. As a result, we get
\begin{align}\label{eq:RLditau}
& \R_L[\partial_i \tau^{\mu\nu}](t) = \mathop{\mathrm{FP}}_{B = 0} \int
\frac{d^3\mathbf{x}'}{-4 \pi} \, \vert\mathbf{x}'\vert^{B+\ell-1}
\,\int_1^{+\infty} dz \Bigl\{ - B n'_i \hat{n}'_L \gamma_\ell(z)
\nonumber \\ & \quad + (2 \ell +1) \bigl[ \bigl(\gamma_\ell(z) -
\gamma_{\ell -1}(z)\bigr) \hat{n}'_{iL} + \gamma_{\ell -1}(z)
\bigl(\hat{n}'_{iL} - n'_i\hat{n}'_L\bigr) \bigr] \Bigr\}
\mathcal{M}(\tau^{\mu\nu}_\text{ns})(\mathbf{x}', t - z
\vert\mathbf{x}'\vert/c) .
\end{align}%
The crucial point is that only a pole $ \propto 1/B $, generated by the radial
integration, may give rise to a non-vanishing contribution for a source term
proportional to $ B $ in the limit $ B\rightarrow 0 $. Therefore, we may
restrict the integration domain over the first term in the curly brackets
above to a ball of radius $ R $ centered at the origin without affecting its
value. An immediate consequence is that $ \mathcal{M}(\tau^{\mu\nu}_\text{ns})
$ can be replaced there by its post-Newtonian or near zone value $
\overline{\mathcal{M} (\tau^{\mu\nu}_\text{ns})} $. Moreover, we have $
\overline{\mathcal{M} (\tau^{\mu\nu}_\text{ns})} = \mathcal{M}
(\overline{\tau}^{\mu\nu}_\text{ns}) $ by the matching between the inner and
exterior fields (see paper II).

The second term entering the brackets of Eq.~\eqref{eq:RLditau}, in view of
the relation $ (d/dz)^2 \bigl(\gamma_{\ell + 1}(z)\bigr) = (2 \ell + 1)(2 \ell
+ 3) \bigl(\gamma_{\ell-1}(z) - \gamma_\ell(z)\bigr) $, is nothing but the
second derivative of $ \gamma_{\ell +1}(z) $ up to a constant factor. This
derivative is handled by means of two successive integrations by parts, and
the resulting derivatives that bear on $ \mathcal{M}(\tau^{\mu\nu}_\text{ns})
$ are transformed like
\begin{equation}
\left(\frac{\partial}{\partial z} \right)^2
\mathcal{M}(\tau^{\mu\nu}_\text{ns})(\mathbf{x}',t - z
\vert\mathbf{x}'\vert/c) = \vert\mathbf{x}'\vert^2
\left(\frac{\partial}{c \partial t}\right)^2
\mathcal{M}(\tau^{\mu\nu}_\text{ns})(\mathbf{x}',t - z
\vert\mathbf{x}'\vert/c) .
\end{equation}%
Taking also into account the fact that $ \hat{n}'_{i L} - n'_i
\hat{n}'_{L} = - \ell \delta_{i\langle i_\ell} \hat{n}'_{L-1\rangle}/(2
\ell +1) $, this yields
\begin{align}
& \R_{L}[\partial_i \tau^{\mu\nu}] + \ell \, \delta_{i\langle i_\ell}
\R_{L-1\rangle}[\tau^{\mu\nu}] + \frac{1}{2 \ell + 3} \left(\frac{d}{c d
t}\right)^2 \R_{i L}[\tau^{\mu\nu}] \nonumber \\ & \quad = -
\mathop{\mathrm{FP}}_{B = 0} B \int \frac{d^3\mathbf{x}'}{-4\pi} \,
\vert\mathbf{x}'\vert^{B-2} x'_{i}\hat{x}'_L \,\int_1^{+\infty} dz \,
\gamma_\ell(z)
\mathcal{M}(\overline{\tau}^{\mu\nu}_\text{ns})(\mathbf{x'},t - z
\vert\mathbf{x}'\vert/c) , \label{eq:R_di_tau}
\end{align}%
which may be used to eliminate $ \R_L[ \partial_i \tau^{\mu\nu}] $ from the
anti-symmetric wave $ \Hom[\partial_i \tau^{\mu\nu}] $.

At this level, we note that at any finite post-Newtonian order, the structure
of $\mathcal{M}(\overline{\tau}^{\mu\nu}_\text{ns})$ is such (see paper II)
that it yields a radial integral in the right-hand side of
Eq.~\eqref{eq:R_di_tau}, that is decomposed into a sum of terms of the type $
\int_R^{+\infty} dr' r'^{B+q} $ with $ q \in \mathbb{Z} $. Now, a useful
lemma, already employed in Refs.~\cite{B95, B98mult} and paper II, states that
$ \int_0^{+\infty} dr' r'^{B+q} = 0$ by analytic continuation in $B$. This
implies that the right-hand side of Eq.~\eqref{eq:R_di_tau} will give zero if
we make the substitution $ \tau^{\mu\nu}_\text{ns} \rightarrow
\tau^{\mu\nu}_\text{s} $, which means that $\tau^{\mu\nu}_\text{ns}$ can be
replaced there by the complete pseudo tensor, $ \tau^{\mu\nu} =
\tau^{\mu\nu}_\text{ns} + \tau^{\mu\nu}_\text{s} $. Our desired commutator,
\textit{i.e.} the left-hand side of Eq.~\eqref{eq:commutator2}, follows
immediately from Eqs.~\eqref{eq:H} and \eqref{eq:R_di_tau}. We find
\begin{equation}\label{eq:comm1}
\bigl[\partial_i, \,\Hom \bigr] [\tau^{\mu\nu}] = \sum_{\ell = 0}^{+
\infty} \frac{(-1)^\ell}{\ell!} \,\hat{\partial}_L \left\{
\frac{\Rs_L^{i}[\tau^{\mu\nu}]( t - r/c ) - \Rs_L^{i}[\tau^{\mu\nu}]( t
+ r/c )}{2r} \right\} ,
\end{equation}%
where
\begin{equation}\label{eq:comm1S}
\Rs_L^{i}[\tau^{\mu\nu}] = \mathop{\mathrm{FP}}_{B = 0} B \int
\frac{d^3\mathbf{x}'}{-4\pi} \, \vert\mathbf{x}'\vert^{B-2}
x'_{i}\hat{x}'_L \,\int_1^{+\infty} dz \, \gamma_\ell(z)
\mathop{\mathcal{M}(\overline{\tau}^{\mu\nu})}(\mathbf{x'},t - z
\vert\mathbf{x}'\vert/c) .
\end{equation}%

In the last stage of our proof, we shall focus on the other commutator in the
right-hand-side of Eq.~\eqref{eq:commutator2}, which is readily obtained by
integration by parts as
\begin{equation}
\bigl[\partial_i, \overline{\Box_\mathrm{R}^{-1}}\bigr] [\tau^{\mu\nu}]
= \sum_{k=0}^{+\infty} \frac{(-1)^k}{k!} \left( \frac{d}{c d t}
\right)^k \mathop{\mathrm{FP}}_{B = 0} B \,\int
\frac{d^3\mathbf{x}'}{-4\pi} \, \vert\mathbf{x}'\vert^{B-2} x'^i
\vert\mathbf{x} - \mathbf{x}'\vert^{k-1}
\overline{\tau}^{\mu\nu}(\mathbf{x}',t) .
\end{equation}%
We shall put it under a form where it is immediate to see that it is the
opposite of $ \bigl[\partial_i, \Hom\bigr] $, which will complete our
demonstration. Due to the presence of a $ B $ factor, the only
contributing terms are those that generate poles in the variable $ B $
due to the divergent behavior of $ \overline{\tau}^{\mu\nu} $ when $
\vert\mathbf{x}' \vert \to + \infty$. Thus, the integral depends only on
the exterior domain, which can be chosen to lie outside a sphere of
arbitrary radius $ R $. If $ R $ is large enough, we are allowed to
replace $ \vert\mathbf{x} - \mathbf{x}'\vert^{k-1} $ and $
\overline{\tau}^{\mu\nu} $ by their multipole expansions, so that
\begin{align}
& \mathop{\mathrm{FP}}_{B = 0} B \int \frac{d^3\mathbf{x}'}{-4\pi} \,
\vert\mathbf{x}'\vert^{B-2} x'_i \vert\mathbf{x} -
\mathbf{x}'\vert^{k-1} \,\overline{\tau}^{\mu\nu}(\mathbf{x}',t) =
\nonumber \\ &\qquad\qquad \mathop{\mathrm{FP}}_{B = 0} B
\mathop{\int_{\vert\mathbf{x}'\vert> R}} \frac{d^3\mathbf{x}'}{-4\pi} \,
\vert\mathbf{x}'\vert^{B-2} x'_i \,\mathcal{M}(\vert\mathbf{x} -
\mathbf{x}'\vert^{k-1})
\mathcal{M}(\overline{\tau}^{\mu\nu})(\mathbf{x}',t).
\end{align}%
By virtue of the lemma mentioned above, we can change the integration over the
domain $ \vert\mathbf{x}'\vert> R $ into an integration over the complementary
domain $ \vert\mathbf{x}'\vert< R $ if we also change the sign. On the other
hand, the expression of $ \mathcal{M}(\vert\mathbf{x} - \mathbf{x}'\vert^{s})
$ for $ s \in \mathbb{N} $ is known explicitly. It is derived from the STF
decomposition of the factors $ x'_L $ (see \textit{e.g.} Eq.~(2.2) in
\cite{Th80} or (A20a) in \cite{BD86}) that enter the term of order $ \ell $ in
the series expansion of $ \vert\mathbf{x} - \mathbf{x}'\vert^{s} $ valid for $
\vert\mathbf{x}'\vert < \vert\mathbf{x}\vert $. We recover the same formula as
Eq.~\eqref{eq:multipole_r2k} but for $ k = s/2 $, namely
\begin{equation} \label{eq:multipole_rs}
\frac{\vert\mathbf{x}-\mathbf{x}'\vert^{s}}{(s+1)!}
=\sum_{\ell=0}^{+\infty}\frac{(-1)^\ell}{\ell!\,(2\ell+1)!!}
\sum_{p=0}^{+\infty}\widetilde{\Delta^{-p}}\left(\hat{x}_L\right)
\widetilde{\Delta^{-\frac{s}{2}+\ell+p}}\left(\hat{x}'_L\right) ,
\end{equation}%
provided that we define, for any integer $s$,
\begin{equation}\label{eq:Delta-sB0}
\widetilde{\Delta^{-\frac{s}{2}}} \left(\hat{x}_L\right) = \frac{(2\ell
+1)!!}{s!! (2\ell+s+1)!!}\,r^s \hat{x}_L ,
\end{equation}%
which generalizes the notation \eqref{eq:Delta-k}. We can transform
\eqref{eq:Delta-sB0}, with the help of the Euler function $
\mathrm{B}\bigl(p, \,q\bigr) = 2 \int_1^{+\infty} dz \, z^{-2p-2q+1}
(z^2-1)^{q-1}$, to
\begin{equation}\label{eq:Delta-sB}
\widetilde{\Delta^{-\frac{s}{2}}} \left(\hat{x}_L\right) = \frac{r^s
\hat{x}_L}{s!} \frac{(-1)^{\ell+1}(2\ell +1)!!}{2^{\ell+1} \ell!}
\,\mathrm{B}\bigl(-\ell-(s+1)/2, \, \ell + 1 \bigr) .
\end{equation}%
The appearance of the Euler function is of particular interest for our
purpose, as it can simply be represented by an integral involving $
\gamma_\ell(z) $. Indeed, we have $ \mathrm{B}\bigl(-\ell-(s+1)/2, \,
\ell +1 \bigr) = 2 \int_1^{+ \infty} dz \, z^{s} (z^2-1)^{\ell}$, which
immediately gives
\begin{equation} \label{eq:Delta-s}
\widetilde{\Delta^{-\frac{s}{2}}} \left(\hat{x}_L\right) = \frac{r^s
\hat{x}_L}{s!} \int_1^{+\infty} dz \, \gamma_\ell(z) \, z^s .
\end{equation}%
Inserting the value of $ \mathcal{M}(\vert\mathbf{x} -
\mathbf{x}'\vert^{k-1}) $ deduced from
Eqs.~\eqref{eq:multipole_rs}--\eqref{eq:Delta-s} into the spatial
integral with the radial variable $ \vert\mathbf{x}'\vert $ ranging from
0 to $ R $ (taking into account the change of sign when going from the
outer to the inner domain), we obtain
\begin{align}
& \bigl[\partial_i, \overline{\Box_\mathrm{R}^{-1}}\bigr]
[\tau^{\mu\nu}] = - \sum_{k,\ell,p\ge 0}
\frac{(-1)^{k+\ell}}{(2\ell+1)!!\ell!} \,\widetilde{\Delta^{-p}}
\left(\hat{x}_L\right) \nonumber \\ & \times \left(\frac{d}{c d t}
\right)^k \! \mathop{\mathrm{FP}}_{B = 0} B \!
\int_{\vert\mathbf{x}'\vert < R} \! \frac{d^3\mathbf{x}'}{-4\pi}
\,\frac{\vert\mathbf{x}'\vert^{B+k - 2\ell - 2p - 3}
x'_{i}\hat{x}'_L}{(k - 2 \ell - 2p - 1)!} \,\int_1^{+\infty} dz \,
\gamma_\ell(z) \, z^{k - 2\ell - 2p - 1}
\mathop{\mathcal{M}(\overline{\tau}^{\mu\nu})}(\mathbf{x}',t).
\end{align}%
Note that only the terms for which $ k \ge 2 \ell + 2 p + 1$ contribute to the
sum over $ k $ since otherwise $ 1/(k - 2\ell - 2p -1)! = 0 $. The divergences
near $ \vert \mathbf{x}'\vert =0 $ can be cured by means of the finite part
regularization FP. We next permute the summation over $ k $ and the
integration on $ z $. We easily recognize, from the summation over $k$, the
Taylor expansion of $
\mathop{\mathcal{M}(\overline{\tau}^{\mu\nu})}(\mathbf{x}',t - z
\vert\mathbf{x}'\vert/c) $ when $c\rightarrow +\infty$. Finally, we obtain the
relation
\begin{align} \label{eq:di_box-1}
& \bigl[\partial_i, \overline{\Box_\mathrm{R}^{-1}}\bigr]
[\tau^{\mu\nu}] = \sum_{\ell,p \ge 0} \frac{(-1)^\ell}{(2\ell+1)!!\ell!}
\,\widetilde{\Delta^{-p}} \left(\hat{x}_L\right) \nonumber \\ & \times
\left(\frac{d}{c d t}\right)^{2\ell+2p+1} \! \! \! \!
\mathop{\mathrm{FP}}_{B = 0} B \int_{\vert\mathbf{x}'\vert \le R}
\frac{d^3\mathbf{x}'}{-4\pi} \, \vert\mathbf{x}'\vert^{B-2}
x'_{i}\hat{x}'_L\,\int_1^{+\infty} dz \, \gamma_\ell (z)
\mathop{\mathcal{M}(\overline{\tau}^{\mu\nu})} (\mathbf{x}',t - z
\vert\mathbf{x}'\vert/c) .
\end{align}
Thanks to the formula~\eqref{eq:antisymmetric_F_wave}, which is valid for any
function, we identify the right-hand side of \eqref{eq:di_box-1} with the
opposite of the commutator found in Eqs.~\eqref{eq:comm1}--\eqref{eq:comm1S},
when it is developed as $c\rightarrow +\infty$. We can therefore conclude that
\begin{equation}
\bigl[\partial_i, \,\mathcal{D}^{-1}\bigr] \equiv \bigl[\partial_i,
\,\Hom\bigr] + \bigl[\partial_i,
\,\overline{\Box_\mathrm{R}^{-1}}\bigr] = 0 ,
\end{equation}
which, as we discussed above, achieves the proof that $
\overline{h}^{\mu\nu} $ satisfies indeed the harmonicity conditions.

\section{Conclusion} \label{sec:conclusion}

In conclusion, this work considers the general post-Newtonian approximation
formalism for isolated, slow-moving matter systems. Our motivation is to
justify the practical 3.5PN calculations of radiation reaction achieved in
paper I (Ref.~\cite{NB05}). Our study follows from the previous investigation
of paper II, in which a general scheme was proposed for the formal iteration
of the post-Newtonian series in the source's near zone up to any
post-Newtonian order. There, the expression comprised: (i) a particular
solution to the flat wave equation, given in terms of a (well-defined)
regularized solution, made with a finite part FP, to the iterated Poisson
equation for a non-compact source, and (ii) a homogeneous solution, defined as
a specific multipolar anti-symmetric wave, associated with radiation reaction
effects. Here, we present an alternative, complementary expression for the
gravitational field in the source's near-zone within the post-Newtonian
approximation; notably, as in paper II, the expression incorporates fully the
effects of the radiation reaction. It involves both particular and homogeneous
solutions. In contrast to paper II, however, the particular solution in this
instance represents a formal expansion of the retardations of the standard
retarded integral, where each term has been suitably regularized using the FP
procedure. The homogeneous second term, given in the form of a multipolar
anti-symmetric wave, describes the non-linear contribution in the radiation
reaction force of gravitational wave tails beginning at 4PN order $\sim
1/c^8$. The ``ordinary'' radiation reaction effects are then contained in the
odd-parity part of the retardations of the particular, inhomogeneous solution.

Consequently, as shown in Ref.~\cite{NB05}, the radiation reaction
effects in the local dynamics of an isolated source up to and including
the 3.5PN order $\sim 1/c^7$ can be determined solely by considering the
odd-parity part of the retardations in the standard retarded integral,
following the appropriate regularization based on a FP prescription.

Finally and complementary to the proof given in paper II, we verify that
our post-Newtonian solution is divergenceless by demonstrating that the
harmonic-coordinate condition is satisfied. Hence, the new expression is
indeed a solution of the Einstein field equations.

Together with useful technical relationships given throughout this
paper, the new form for the metric in the near-zone enables significant
insights into the character of the gravitational radiation reaction and
contribution of tails for isolated matter systems in the post-Newtonian
expansion of general relativity.

\acknowledgments

S.N. gratefully acknowledges the Entente Cordiale Scholarship and Robert Blair
Fellowship (Corporation of London) for financial support during this work.    

\bibliography{BFN05}

\end{document}